\documentclass[aps,showpacs]{revtex4}
\usepackage{graphicx}
\def\bc{\begin{center}}
\def\ec{\end{center}}

\def\beq{\begin{equation}}
\def\eeq{\end{equation}}

\begin{document}

\title{Photonic spectral density of coupled microwave cavities
}

\author{K. Ziegler}
\affiliation{Institut f\"ur Physik, Universit\"at Augsburg, D-86135 Augsburg, Germany}
\date{\today}

\begin{abstract}
We study a pair of anharmonic microwave cavities that is connected by an optical fiber.
The photonic spectral density characterizes the evolution of the coupled cavities
after the system has been prepared in a Fock or N00N state. We evaluate the photonic spectral density
within the recursive projection method and find that the anharmonicity has no substantial effect on the
spectral properties. In all cases the photonic spectral density has a Gaussian envelope for large
photon numbers.
\end{abstract}
\pacs{42.50.Ar, 42.50.Pq, 42.50.Ct}

\maketitle

\section{Introduction}

\begin{figure}
\begin{center}
\includegraphics[width=7cm,height=2.5cm]{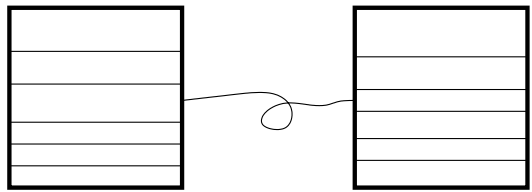}
\caption{
Two microwave cavities coupled by an optical fiber. The energy levels
are indicated as those of anharmonic cavities.
}
\label{cavities}
\end{center}
\end{figure}

The experimental preparation of Fock states in a microwave cavity \cite{brune08,wang08} is a crucial
step towards a systematic study of correlated many-body systems with photonic states. 
Once a Fock state has been prepared, we can change the parameters of the system 
and study the evolution of the Fock state within the Hilbert space that is provided by the eigenstates of the 
Hamiltonian of the new system. This can be realized, for instance, in an 
experiment with two microwave cavities. The system is prepared such that one cavity has $N$ photons, the other cavity has
no photon. Then the initial Fock state $|N,0\rangle$ is assumed to be the groundstate of a Hamiltonian $H_0$. 
At the time $t=0$ the Hamiltonian $H_0$ is switched to the Hamiltonian $H$. This problem has been studied intensively
for atomic systems, using the Hubbard model and related models \cite{kollath07,rigol07,manmana07,eckstein08,moeckel08,kollar08}.
In the case of two microwave cavities 
this switch can be realized by connecting the two isolated cavities with an optical fiber. Then $H_0$
is the Hamiltonian of the isolated cavities, and $H$ is the Hamiltonian of the connected cavities. 
This type of system, including atomic degress of freedom, was studied recently within a Hartree-Fock approximation
\cite{ji09}.
In this paper we shall study the dynamics of photons which tunnel between two microwave
cavities through an optical fiber (cf. Fig. \ref{cavities}). The effect of atoms inside the cavities
is approximated by an anharmonic photonic term. A central question is the effect of this
anharmonicity on the evolution and whether or not the spectral properties depend on the initial 
conditions. 


\section{Dynamics of an isolated system}

We consider a system which is isolated from the environment. In terms of photonic states this can be realized by
an ideal microwave cavity. Furthermore, we assume that the system lives in an $N+1$-dimensional Hilbert space. 
With the initial state $|\Psi_0\rangle$  
we obtain for the time evolution of the state
\[
|\Psi_t\rangle=e^{-iHt}|\Psi_0\rangle
\]
or the evolution of the return probability $|\langle\Psi_0|\Psi_t\rangle|^2$ with the amplitude
\beq
\langle\Psi_0|\Psi_t\rangle =\langle\Psi_0|e^{-iHt}|\Psi_0\rangle 
\ .
\label{overlap}
\eeq
A Laplace transformation relates the return amplitude with the resolvent through the identity
\[
\langle\Psi_0|\Psi_t\rangle 
=\int_\Gamma \langle\Psi_0|(z-H)^{-1}|\Psi_0\rangle e^{-izt}dz
\ ,
\]
where the contour $\Gamma$ encloses all the eigenvalues $E_j$ ($j=0,1,...,N$) of $H$.
With  the corresponding eigenstates $|E_j\rangle$ the spectral representation of the resolvent provides a rational function:
\beq
\langle\Psi_0|(z-H)^{-1}|\Psi_0\rangle=\sum_{j=0}^N\frac{\langle\Psi_0|E_j\rangle}{z-E_j}
=\frac{P_N(z)}{Q_{N+1}(z)}
\ ,
\label{resolvent2}
\eeq
where $P_N(z)$, $Q_{N+1}(z)$ are polynomials in $z$ of order $N$, $N+1$, respectively. 
These polynomials can be evaluated by the 
recursive projection method (RPM) \cite{ziegler10a}. This method is based on a systematic expansion of
the resolvent $\langle \Psi_0|(z-H)^{-1}|\Psi_0\rangle$, starting from the initial state $|\Psi_0\rangle$.
The method can also be understood as a random walk in Hilbert space. However, there is a crucial difference
in comparison with the conventional random walk because the RPM has a prefered direction: 
The walk of the RPM visits a subspace ${\cal H}_{2j}$ only once and never returns to it.
A step  from ${\cal H}_{2j}$ to ${\cal H}_{2j+2}$ 
is defined by the Hamiltonian $H$ in such a way that 
${\cal H}_{2j+2}$ is created by acting $H$ on ${\cal H}_{2j}$. 
In terms of $N$ photons, distributed over two microwave cavities, these subspaces are given by the
basis $\{|N-j,j\rangle,|j,N-j\rangle\}$, and the a step is an exchange of a single 
photon. Thus, the walk follows a path with increasing
numbers $j$. This is the main advantage of the RPM that allows us to calculate the resolvent 
$\langle \Psi_0|(z-H)^{-1}|\Psi_0\rangle$ on a $N+1$-dimensional Hilbert space exactly. In case of the
photons in two microwave cavities the RPM is efficient for values of $N$ up to the order of $\approx 10^4$.

The expression in Eq. (\ref{resolvent2}) suggests the introduction of the photonic spectral density (PSD) 
$\rho_\epsilon(E)$ as the imaginary part of the resolvent:
\beq
\rho_\epsilon(E)=Im \langle \Psi_0|(E-i\epsilon-H)^{-1}|\Psi_0\rangle
= \epsilon\sum_{j=0}^N\frac{|\langle\Psi_0|E_j\rangle|^2}{\epsilon^2+(E-E_j)^2}
\ .
\label{spectrald}
\eeq
The amplitude of the return probability then reads as the Fourier transform of the PSD
\beq
\langle\Psi_0|\Psi_t\rangle
=\lim_{\epsilon\to 0}\int\rho_\epsilon(E)e^{-iEt}dE
\ .
\label{fourier}
\eeq
Moreover, the overlap $\langle\Psi_0|E_j\rangle$ between the initial state and the eigenstates of the Hamiltonian $H$ 
are directly obtained at the poles of the PSD. 
In the following the PSD shall be evaluated with the RPM, using Fock states $|0,N\rangle$, $|N,0\rangle$ and superpositions
of these states.

\section{single cavity}

We begin with the simplest case, a harmonic cavity which has a harmonic oscillator spectrum
\[
H_0=\omega_0 a^\dagger a
\]
with cavity mode $\omega_0$ and the ceation (annihilation) operator $a^\dagger$ ($a$) of a photon.
The eigenstates are Fock states with $N$ photons $|N\rangle$ ($N=0,1,...$) and energy 
$E_N=\langle N|H_0|N\rangle=\omega_0 N$.

\subsection{Single atom: Jaynes-Cummings model}

An anharmonicity can be created by adding atoms to the cavity which interact with the photons.
After introducing a single two-level atom to the cavity we can describe the absorption and emission
of photons by the atom approximately with the Jaynes-Cummings model \cite{jaynes63}, whose Hamiltonian
reads
\[
H_{JC}= \omega_0 a^\dagger a +(\omega_0+\Delta) c^\dagger c
-g(a^\dagger c+c^\dagger a)
\ .
\]
$\Delta$ is the detuning between the atomic excitation energy and the photon energy, and
$c^\dagger$ ($c$) is the creation (annihilation) operator of the atomic excitation.
The eigenvalues of this Hamiltonian are \cite{jaynes63}
\beq
E_{N\pm}=\omega_0(N+1/2)\pm\sqrt{\Delta^2+4g^2(N+1)}
\label{jc_spectrum}
\eeq
with eigenstates
\[
|N,\pm\rangle=\frac{1}{\sqrt{2}}[\pm |N,1\rangle+|N+1,0\rangle]
\ .
\]
Thus, the eigenstates are not Fock states 
but a superposition of 
two Fock states, one with $N$ photons and the atom in the groundstate
and one with $N-1$ photons and the atom in the excited state. The transition
amplitude from the initial Fock state $\langle N,0|\Psi_t\rangle$ is a simple oscillation 
because only $|N-1,\pm\rangle$ contributes. The superposition of many Fock states can lead
to a more complex behavior, such as a collapse and revival dynamics \cite{puri85,buzek89}.
Ignoring the atomic degrees of freedom, the effect of the atom on the photonic spectrum is
the additional square-root term in Eq. (\ref{jc_spectrum}).

\subsection{Many atoms: anharmonic cavity}

The presence of many atoms in the cavity is quite complex due the presence of the atomic degrees of 
freedom. Therefore, we apply a simplified picture by ignoring the atomic degrees of freedom and 
describing their effect on the photons as an anharmonicity in the photonic spectrum: 
\beq
\omega_0 a^\dagger a\to f(a^\dagger a)
\ .
\label{general1}
\eeq
Although not much is known for a realistic change of the photonic spectrum due to atoms, a possible
anharmonic term is the interaction of the Hubbard model
\beq
f(a^\dagger a)=\omega_0 a^\dagger a+U(a^\dagger a)^2
\ .
\label{hubbarda}
\eeq
This quadratic anharmonicity probably overestimates the contribution of the atoms in the
cavity. A more realistic case might be a $\sin^2$ anharmonic term:
\beq
f(a^\dagger a)=\omega_0 a^\dagger a+U\sin^2(\alpha a^\dagger a)
\ .
\label{anharm1}
\eeq
These two examples present a basis for analyzing the generic evolutionary properties of coupled
anharmonic microwave cavities.

\section{Two coupled cavities}

The Hamiltonian of two coupled cavities
\beq
H=-J(a_1^\dagger a_2+a_2^\dagger a_1) + f(a^\dagger_1 a_1)+f(a^\dagger_2 a_2)
\label{Hamiltonian2}
\eeq
describes the tunneling of photons through the optical fiber with rate $J$.
The properties of the two cavities are assumed to be equal and given by $f(a^\dagger_j a_j)$.
While Fock states are eigenstates of the disconnected cavities (i.e. for $H_0=f(a^\dagger_1 a_1)+f(a^\dagger_2 a_2)$),
this is not the case for the coupled system. This implies that a Fock state as the initial state undergoes an 
evolution by visiting the eigenstates of $H$.  The average energy ${\bar E}$ is conserved, since the evolution of an 
isolated system is a unitary transformation: 
\beq
{\bar E}=\langle\Psi_t|H|\Psi_t\rangle=\langle\Psi_0|H|\Psi_0\rangle=f(N-k)+f(k)
\ ,
\label{energy}
\eeq
where $k$ and $N-k$ are the number of photons in the cavities for the initial state.

The coupled microwave cavities with the two Fock states $|N,0\rangle$, $|0,N\rangle$ as initial states
can be treated within the RPM.  This is a generalization of the RPM of Ref. \cite{ziegler10a} for a single Fock state 
$|N,0\rangle$ to a two-dimensional Fock basis. Assuming that $N$ is even, all projected spaces ${\cal H}_{2j}$ 
are two dimensional and spanned by $\{|N-j,j\rangle , |j,N-j\rangle\}$ ($j=0,...,N/2$). 
This leads to a recurrence relation which can accommodate the Fock states $|N,0\rangle$, $|0,N\rangle$.
In terms of Pauli matrices this reads 
$\{\sigma_j\}_{j=0,...,3}$
\beq
g_{k+1}=a_{k+1}\sigma_0+b_{k+1}\sigma_1, \ \ 
g_0=\frac{1}{z-2f(N/2)}\sigma_0 \ \ \ (k=0,1,...,N/2-1)
\label{2d_recursion}
\eeq
with coefficients
\beq
a_{k+1}=\frac{z-{\tilde f}_{k+1}-J^2a_k(N/2+k+1)(N/2-k)}
{\left[z-{\tilde f}_{k+1}-J^2a_k(N/2+k+1)(N/2-k)\right]^2-J^4b_k^2(N/2+k+1)^2(N/2-k)^2}
\label{coeff_a}
\eeq
\beq
b_{k+1}=\frac{J^2b_k(N/2+k+1)(N/2-k)}
{\left[z-{\tilde f}_{k+1}-J^2a_k(N/2+k+1)(N/2-k)\right]^2-J^4b_k^2(N/2+k+1)^2(N/2-k)^2}
\label{coeff_b}
\eeq
and
\[
{\tilde f}_{k+1}=f(N/2+k+1)+f(N/2-k-1)
\ .
\]
The initial conditions for the coefficients are
\[
a_0=b_0=\frac{1}{z-2f(N/2)}
\]
and the termination of the iteration gives
\beq
g_{N/2}=a_{N/2}\sigma_0+b_{N/2}\sigma_1
\ .
\label{termination}
\eeq
Here the Pauli-matrix structure is based on the two Fock states $|0,N\rangle$ and $|N,0\rangle$.
It should be noticed that the cavity mode $\omega_0$ appears as $\omega_0 N$ in ${\tilde f}_k$.
Thus, this term leads only to a global shift of the energy $z\to z-\omega_0N$.

From the iteration of the recurrence relations (\ref{coeff_a}), (\ref{coeff_b}) we obtain the coefficients 
$a_{N/2}$ and $b_{N/2}$ of $g_{N/2}$. This allows us to evaluate the polynomials $P_N(z)$, $Q_{N+1}(z)$
of the resolvent in Eq. (\ref{resolvent2}) for the two Fock states $|N,0\rangle$, $|0,N\rangle$
as well as for the entangled even and antisymmetric N00N states
\beq
|N00N\rangle=\frac{1}{\sqrt{2}}\left(|N,0\rangle +|0,N\rangle\right) , \ \ \
|N00N\rangle'=\frac{1}{\sqrt{2}}\left(|N,0\rangle -|0,N\rangle\right)
\label{noon}
\eeq
as initial states. Eq. (\ref{termination}) implies for a Fock state as initial state
\beq
\langle N,0|(z-H)^{-1}|N,0\rangle = \langle 0,N|(z-H)^{-1}|0,N\rangle = a_{N/2}
\ ,
\label{fock1}
\eeq
for the symmetric N00N state as initial state
\beq
\langle N00N|(z-H)^{-1}|N00N\rangle = a_{N/2}+b_{N/2}
\ ,
\label{noon1}
\eeq
and for the antisymmetric N00N state as initial state
\beq
'\langle N00N|(z-H)^{-1}|N00N\rangle' = a_{N/2}-b_{N/2}
\ .
\label{noon2}
\eeq

\section{results}

\begin{figure}[t]
\begin{center}
\includegraphics[width=12cm,height=7cm]{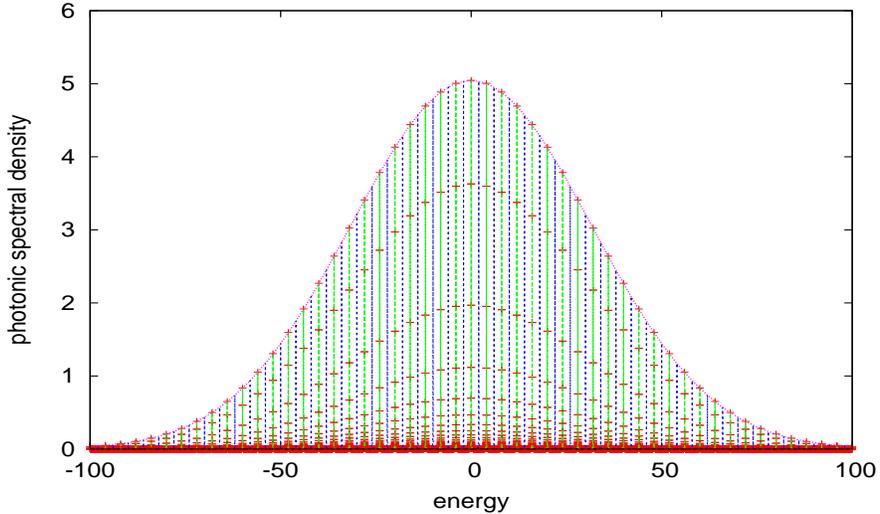}
\caption{
Photonic spectral function for 1000 photons in harmonic cavities with $J=1$, $\epsilon=0.01$, using symmetric and antisymmetric N00N states.
The states with even parity are marked by crosses.}
\label{fig:0}
\end{center}
\end{figure}

We start with tunneling between two harmonic cavities, where $f(N)=\omega_0 N$.  
In this case the denominator of $a_{N/2}$, $b_{N/2}$ factorizes: 
\beq
Q_{N+1}(z)=z\prod_{k=1}^{N/2}[(2kJ)^2-z^2]  
\ .
\label{coupled_harmonic}
\eeq
In this expression we have dropped the global shift of the energy $z$ by the harmonic term $\omega_0 N$.
The distribution of the peaks of the PSD has a Gaussian envelope for both, the Fock and the N00N state
as initial state (cf. Fig. \ref{fig:0}). This can be directly explained by calculating the overlap between the 
eigenstates of the tunneling term $|N,k\rangle_t$ ($k=0,..,N$) and the Fock state:
\beq
\langle 0,N|N,k\rangle_t=2^{-N/2}{N\choose k}^{1/2}
\ .
\label{binomial}
\eeq
Thus there is a binomial distribution for the overlap $|\langle 0,N|N,k\rangle_t|^2$
which becomes a Gaussian distribution for large $N$.
The Gaussian result indicates a kind of Central Limit Theorem for 
independent photons. Such a behavior was also found previously
for the free expansion of bosons from an initial Fock states \cite{cramer08}. 

For the case of anharmonic cavities the analytic expressions of the polynomials $P_N(z)$, $Q_{N+1}(z)$
are quite lengthy, and it is unknown if they can be written in a simple form such as Eq. (\ref{coupled_harmonic}). 
Therefore, it is more convenient to represent the results of the iterations for 1000 photons as plots of the PSD, 
using the expression of Eq. (\ref{spectrald}). The contribution of the harmonic term $\hbar\omega_0N$
appears only as a global energy shift and is ignored in the subsequent discussion.

\begin{figure}[t]
\begin{center}
\includegraphics[width=12cm,height=7cm]{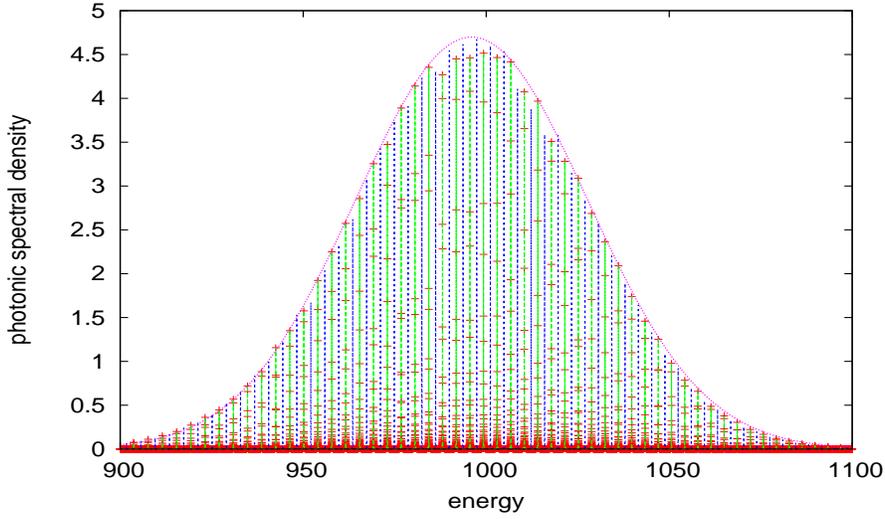}
\caption{
Photonic spectral function for 1000 photons in Hubbard cavities with $J=1$ and 
$U=0.001$. 
The energy of the system is ${\bar E}=1000$. 
}
\label{fig:h1}
\end{center}
\end{figure}

\begin{figure}[t]
\begin{center}
\includegraphics[width=12cm,height=7cm]{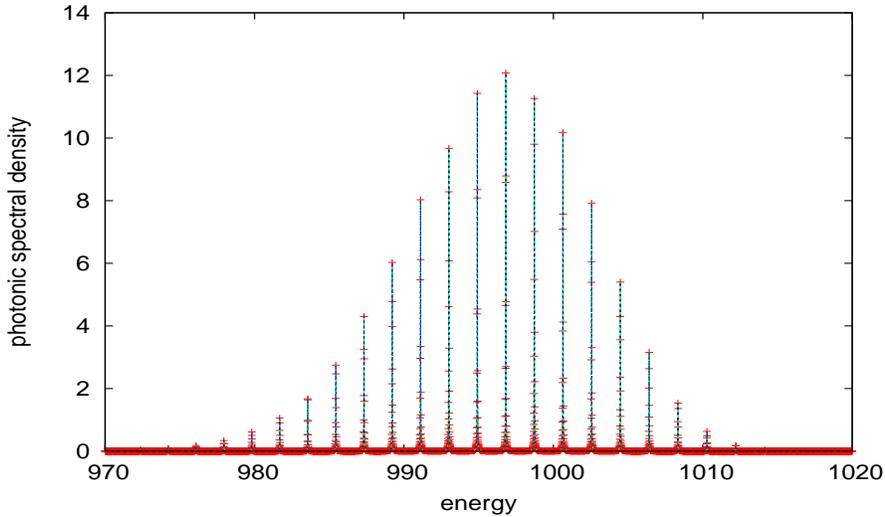}
\caption{
Photonic spectral function for 1000 photons in Hubbard cavities with weaker tunneling rate $J=0.2$ and
$U=0.001$. The PSD is much narrower than for larger tunneling rates.
}
\label{fig:h2}
\end{center}
\end{figure}

\begin{figure}[t]
\begin{center}
\includegraphics[width=12cm,height=7cm]{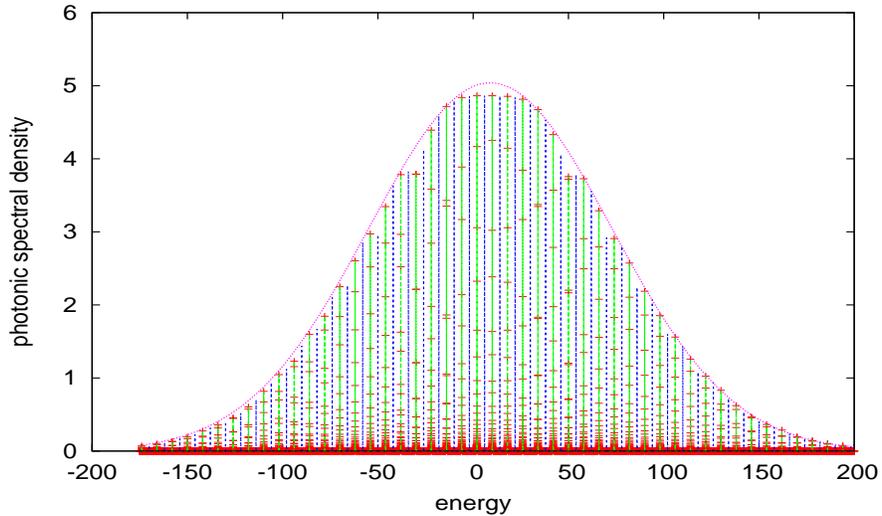}
\caption{
Photonic spectral function for 1000 photons in $\sin^2$ cavities with $U=10$ and $J=4$, $\epsilon=0.01$, $a=3.7119$.
The energy of the system is ${\bar E}\approx 10$.} 
\label{fig:s1}
\end{center}
\end{figure}

\begin{figure}[t]
\begin{center}
\includegraphics[width=12cm,height=7cm]{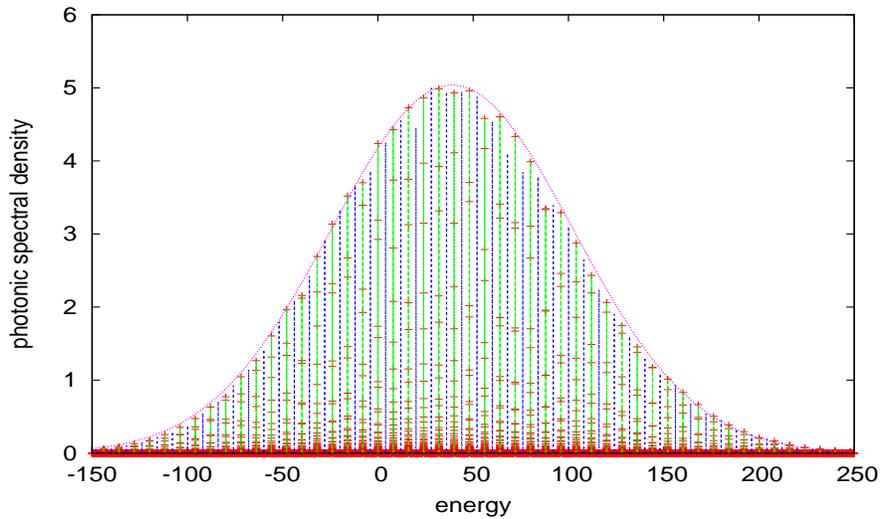}
\caption{
Photonic spectral function for 1000 photons in $\sin^2$ cavities with $U=40$, the other parameters are the same as in the previous figure.
The energy of the system is ${\bar E}\approx 40$.}
\label{fig:s2}
\end{center}
\end{figure}

The PSD is shown for harmonic cavities (Fig. \ref{fig:0})
as well as for anharmonic cavities with a Hubbard anharmonicity (Figs. \ref{fig:h1}, \ref{fig:h2})
and with a $\sin^2$ anharmonicity in Figs. \ref{fig:s1}, \ref{fig:s2}.
The initial states are the two N00N states defined in Eq. (\ref{noon}).   
This allows us also to determine the PDS for the Fock states as initial states by linear combination.
A characteristic feature of all the plots is the Gaussian envelope of the PDS, centered at
the energy of the system in the absence of tunneling ${\bar E}=f(N)$. The tunneling rate $J$ determines the 
width (cf. Fig. \ref{fig:h2}).
Moreover, the spikes of the PSD depend on the initial state, where the symmetric and the antisymmetric N00N state misses 
every other spike. This is also the case for harmonic and for anharmonic cavities. The reason is that the 
eigenstates appear either with even or with odd parity.  Since the symmetric N00N state has even parity, its 
overlap with eigenstates of odd parity vanishes, and vice versa. 
The plots of the PSD in Figs. \ref{fig:h1}-\ref{fig:s2} indicate that every other state has even (odd) 
parity for all considered anharmonicities . The fact that states with odd and 
even parity are alternating results in a suppression of level crossing.

\section{Discussion and conclusions}


The overlap distribution function of the eigenfunctions with the initial state $p_j=|\langle\Psi_0|E_j\rangle|^2$ 
has been used in the literature to study various physical quantities, such as the entropy or the time evolution of the return 
probability of finite isolated quantum systems, e.g., the Bose-Hubbard model. 
This quantity is also useful to characterize the thermalization \cite{roux09,rigol10}. For the Bose-Hubbard model 
this distribution function has a rich structure.  For some parameter values it shows an exponentially 
decaying envelope though. In our plots the distribution function $p_j=|\langle\Psi_0|E_j\rangle|^2$ is 
the envelope of the PSD, multiplied by $\epsilon$ (cf. Eq. (\ref{spectrald})).

Contrary to the exponential envelope of the Bose-Hubbard model in Refs. \cite{roux09,rigol10} we find a 
Gaussian envelope for the coupled cavities. For smaller systems we saw deviations from this behavior, at least for
the Hubbard anhamonicity \cite{ziegler10a}. 
This structure is not very sensitive to the model parameter, as shown in Figs. \ref{fig:h1}-\ref{fig:s2}, 
indicating a random-matrix type of behavior, similar to that found in other few-body systems \cite{porter,mehta,brody81}. 
In this context it is interesting to notice that a mean-field (Hartree-Fock) approximation of our model
in Eq. (\ref{Hamiltonian2}) for a Hubbard anharmonicity gives a relatively simple behavior \cite{milburn97}:
The photonic occupation $N_t$ of one cavity is a periodic function of time $t$ and reads
\[
N_t=\frac{N}{2}[1+cn(Jt|N^2/N_c^2)]
\ ,
\]
where the critical number of atoms is $N_c=J/U$. The Jacobian elliptic functions $cn(Jt|N^2/N_c^2)$ 
is periodic in the first argument $Jt$ with equidistant frequencies
\[
E_n=\nu n ,\ \ \
\nu=\frac{\pi J}{2 K} , \ \ \ K=\int_0^{\pi/2}\frac{1}{\sqrt{1-N^2\sin^2(\Theta)/N_c^2}}d\Theta
\ .
\]
In particular, at the critical point $N=N_c$ it decays like ${\rm sech}(Jt)=1/\cosh(Jt)$. Such a 
behavior is quite different from our results of the Hubbard anharmonicity in Figs. \ref{fig:h1}, \ref{fig:h2}.

The PSD reveals details of the evolution of photons which are tunneling between two microwave cavities.
The fact that eigenstates appear pairwise with even and odd parity implies a strong dependence on the parity 
of the initial state. 
Thus the PSD provides a signature that the evolution of the return probablity $|\langle\Psi_0|\Psi_t\rangle|^2$
depends on the choice of the initial state. 
This can be used to filter states according to parity in the evolution by choosing an appropriate initial state. 

The PSD is invariant with respect to the harmonic frequency of the cavities, except for a global energy shift. 
This reflects an important  universality of the PSD that allows us to separate the harmonic from the anharmonic 
properties of the cavities. 
The behavior of the PSD for harmonic and anharmonic cavities is characterized by a Gaussian envelope, implying that 
the overlap of the eigenfunctions with the initial state $p_j=|\langle\Psi_0|E_j\rangle|^2$ has a discrete Gaussian distribution. 
This simple feature is valid for sufficiently large photon numbers with deviations for smaller photon numbers though.
Comparing the results of the RPM with earlier results of a Hartree-Fock approximation indicates that the
latter is not a good approximation of the PSD, since the energy levels are very different. This does not rule
out that Hartree-Fock approximation can be a useful tool for expectation values of some physical quantities.

\end{document}